\documentclass[conference]{IEEEtran}
\IEEEoverridecommandlockouts
\usepackage{cite}
\usepackage{amsmath,amssymb,amsfonts}
\usepackage{algorithmic}
\usepackage{graphicx}
\usepackage{textcomp}
\usepackage{xcolor}
\usepackage{booktabs}
\usepackage{multirow}
\usepackage{amsmath}
\usepackage{subfigure}
\usepackage{pifont}
\usepackage{xspace}
\usepackage{algorithm}
\usepackage{algorithmic}
\usepackage[square,sort,comma,numbers]{natbib}
\newcommand{\eg}{\emph{e.g.}\xspace}
\newcommand{\ie}{\emph{i.e.}\xspace}

\def\BibTeX{{\rm B\kern-.05em{\sc i\kern-.025em b}\kern-.08em
    T\kern-.1667em\lower.7ex\hbox{E}\kern-.125emX}}
\begin{document}

\title{{\textsc{ReAssigner}}: A Plug-and-Play Virtual Machine Scheduling Intensifier for Heterogeneous Requests
\thanks{This work was supported in part by the National Key R$\&$D Program of China (No. 2022YFA1003900), NSFC (No. 12071145), Huawei Project and the Open Research Projects of Zhejiang Lab (NO.2021KE0AB03). This paper is funded by Shanghai Trusted Industry Internet Software Collaborative Innovation Center. Corresponding author: Xiangfeng Wang.}
}

\author{\IEEEauthorblockN{Haochuan Cui\IEEEauthorrefmark{1},
Junjie Sheng\IEEEauthorrefmark{1},
Bo Jin\IEEEauthorrefmark{1},
Yiqiu Hu\IEEEauthorrefmark{2},
Li Su\IEEEauthorrefmark{2},
Lei Zhu\IEEEauthorrefmark{3},
Wenli Zhou\IEEEauthorrefmark{2},
Xiangfeng Wang\IEEEauthorrefmark{1}}
\IEEEauthorblockA{\IEEEauthorrefmark{1}School of Computer Science and Technology, East China Normal University, Shanghai, China 200062}
\IEEEauthorblockA{\IEEEauthorrefmark{2}Algorithm Innovation Lab. Huawei Cloud Computing Technologies Co., Xi'an, Shaanxi, China 710049 }
\IEEEauthorblockA{\IEEEauthorrefmark{3}Computing Service Product Dept. Huawei Cloud Computing Technologies Co., Xi'an, Shaanxi, China 710049}
}
\IEEEoverridecommandlockouts
\IEEEpubid{\makebox[\columnwidth]{978-1-6654-8045-1/22/\$31.00~\copyright2022 IEEE\hfill} \hspace{\columnsep}\makebox[\columnwidth]{ }}

\maketitle
\IEEEpubidadjcol
\begin{abstract}
With the rapid development of cloud computing, virtual machine scheduling has become one of the most important but challenging issues for the cloud computing community, especially for practical heterogeneous request sequences.
By analyzing the impact of request heterogeneity on some popular heuristic schedulers, it can be found that existing scheduling algorithms can not handle the request heterogeneity properly and efficiently.
In this paper, a plug-and-play virtual machine scheduling intensifier, called {\em{Resource Assigner}} ({\sc{ReAssigner}}), is proposed to enhance the scheduling efficiency of any given scheduler for heterogeneous requests.
The key idea of {\sc{ReAssigner}} is to pre-assign roles to physical resources and let resources of the same role form a virtual cluster to handle homogeneous requests.
{\sc{ReAssigner}} can cooperate with arbitrary schedulers by restricting their scheduling space to virtual clusters.
With evaluations on the real dataset from Huawei Cloud, the proposed {\sc{ReAssigner}} achieves significant scheduling performance improvement compared with some state-of-the-art scheduling methods.
\end{abstract}

\begin{IEEEkeywords}
cloud computing, scheduling, capacity planning
\end{IEEEkeywords}

\section{Introduction}

Recently, cloud computing has been widely adopted in practical applications.
Many companies, \eg, Netflix and LinkedIn, or customer individuals, usually buy cloud computing resources from cloud service providers to meet their computing demands.
Cloud service providers host a large number of physical machines (PMs) as a resource pool (cluster) to satisfy customers' demands.
They typically generate virtual machines (VMs), which are hosted on PMs, for customers.
The VM scheduling problem is the key issue in cloud computing, \ie, which PM should be employed to generate the corresponding VM for the current request.
The main objective for the cloud service provider is to find a VM scheduling policy that can satisfy as many demands as possible for a given number of PMs.
If the scheduling policy can handle more requests than others, the cloud service providers would require fewer PMs with reduced purchasing costs.
Due to its potential to bring economic profits, the virtual machine scheduling problem has attracted significant interest in industrial areas.

Heuristic scheduling methods are taken as good choices for large cloud service providers.
First-Fit and Best-Fit~\citep{bays1977comparison} have been proposed for many years, with acceptable performance in cloud scheduling.
However, requests often span multiple companies and individuals, causing high heterogeneity~\citep{reiss2012heterogeneity}.
With the variation of customer demands, the cluster typically needs to handle heterogeneous requests.
Directly applying First-Fit and Best-Fit would lead PMs to an unbalanced resource utilization status.
The unbalanced utilization causes some resources, \eg, memory and CPU, difficult to be utilized efficiently with degraded scheduling results.

This paper first analyzes the impact of request heterogeneity on some popular heuristic schedulers based on traces collected from Huawei Cloud.
The structures of the VM requests in the given dataset vary extremely, \eg, the CPU requirements are from 2 Cores to 192 Cores, and the resource preferences from CPU to memory are different.
Informally, requests that prefer more CPU resources than memory are called CPU-Intensive requests, and others are MEM-Intensive requests.
By conducting different settings on heterogeneity, sufficient empirical results show that the mixture of CPU-Intensive requests and MEM-Intensive requests make a critical impact on current heuristic scheduling methods. 

To improve the performance in the environment with heterogeneous requests, this paper further proposes a plug-and-play scheduling intensifier called {\em{Resource Assigner}} ({\sc{ReAssigner}}).
The core idea of {\sc{ReAssigner}} is to pre-assign roles to resources and to let resources of the same role form a virtual cluster to handle homogeneous requests. 
{\sc{ReAssigner}} has two roles for resources and requests: CPU-Intensive and MEM-Intensive.
When resources are assigned as a specific role, they are only allocate-able for the same role requests.
Specifically, {\sc{ReAssigner}} establishes two virtual clusters, and each cluster is initiated with the same resources as the physical cluster.
{\sc{ReAssigner}} consists of three sub-modules: the {\em{assign}} sub-module, the {\em{categorize}} sub-module, and the {\em{unassign}} sub-module to guide the scheduling process.
The {\em{assign}} and {\em{unassign}} sub-modules assign and unassign roles to resources in the virtual clusters.
The {\em{categorize}} sub-module categorizes each incoming requests to a role.
By restricting the scheduler scheduling requests in their corresponding virtual cluster, our {\sc{ReAssigner}} can cooperate with arbitrary schedulers.
The main components of {\sc{ReAssigner}} are {\em{assign}} and {\em{unassign}}, we propose a max-utilization scheme and an adaptive unassignment scheme for their specific design. 
Finally, we evaluate {\sc{ReAssigner}} on real traces of Huawei-East-1 dataset~\citep{VMAgent} with different heuristic schedulers.
Empirical results show that the proposed {\sc{ReAssigner}} can significantly improve many of the existing schedulers on both allocation length and resource waste reduction under various scenarios. 

The main contributions can be summarized as follows: 1). This paper first empirically analyzes the impact of request heterogeneity on the scheduling performances;
2). This paper proposes a novel plug-and-play virtual machine scheduling intensifier, called {\em{Resource Assigner}} to mitigate the heterogeneity influences;
3). Experiments on real traces of Huawei Cloud show that the proposed {\sc{ReAssigner}} can improve both the allocation length and waste reduction performance under different scenarios.

The rest of the paper is organized as follows.
Section~\ref{sec: relatedwork} contains the related works.
Section~\ref{sec: background} introduces the background and preliminary of cloud scheduling. 
Section~\ref{sec: heterogeneity} proposes the heterogeneity impact of cloud scheduling.
Section~\ref{sec: approach} proposes the {\sc{ReAssigner}} and section~\ref{sec: evaluation} shows the experiments. 
In section~\ref{sec: conclusion}, we make the conclusion of our paper.

\section{Related Work}\label{sec: relatedwork}

Virtual machine scheduling plays a crucial role in cloud computing. Many works have been proposed to improve the load balance, allocation length, and robustness~\citep{delgado2015hawk}.
Currently, various cloud service providers have proposed their own schedulers to meet their business needs, \eg, Protean~\citep{hadary2020protean}, Sparrow~\citep{ousterhout2013sparrow}, Apollo~\citep{boutin2014apollo}, and Omega~\citep{schwarzkopf2013omega}. 
With the growing amount of data, data processing workloads and the management of their resource usage becomes increasingly important, especially when the analysis yields meaningful knowledge~\citep{grzegorowski2021cost}~\citep{zdravevski2019cluster}~\citep{scheinert2021potential}.

Heterogeneity has been studied on various datasets. 
Heterogeneity is a trend in the current development of cloud scheduling~\citep{delimitrou2013paragon} because of the diversity of performance requirements~\citep{zhang2018mixheter}. 
\citet{reiss2012heterogeneity} first analyzed the Google Trace data from a sizable multi-purpose cluster and proposed that heterogeneity is the most notable workload characteristic. \citet{amvrosiadis2018diversity} released four new traces from two private and two High-Performance Computing (HPC) clusters, which shows the similarity related to the Google workload ~\citep{reiss2012heterogeneity}. \citet{lu2017imbalance} revealed several important insights about different types of imbalance in the Alibaba cloud. 
However, these papers do not optimize the heterogeneity in the scheduling process, while this paper proposes {\em{Resource Assigner}} to mitigate the heterogeneity influences.
And to our knowledge, this paper is the first work to analyze the Huawei public dataset.

Many heuristic algorithms have been studied for the dynamic virtual machine scheduling problem in the simulation environment. \citet{JAYA} propose to use the JAYA algorithm for searching for optimal placement and minimizing the energy consumption of the data center. 
\citet{HHSA} introduced a hyper-heuristic scheduling algorithm (HHSA) that integrates multiple basic heuristic algorithms to ﬁnd better scheduling solutions for cloud computing systems. \citet{ILPwithML} formulated an integer linear programming (ILP)-based VM allocation method combined with machine learning to minimize energy consumption and data communication. 
\citet{sheng2022learning} uses reinforcement learning to deal with the similar scheduling problem as this paper, to obtain high-quality solutions with the excellent exploration ability of reinforcement learning.
Although these approaches can obtain optimal results on small scales but are still impractical at runtime on large-scale applications.
And meanwhile, these methods do not optimize the heterogeneity as well.

\section{Background and Preliminaries} 
\label{sec: background}
\begin{figure}[htbp]
\centerline{\includegraphics[width=.46\textwidth]{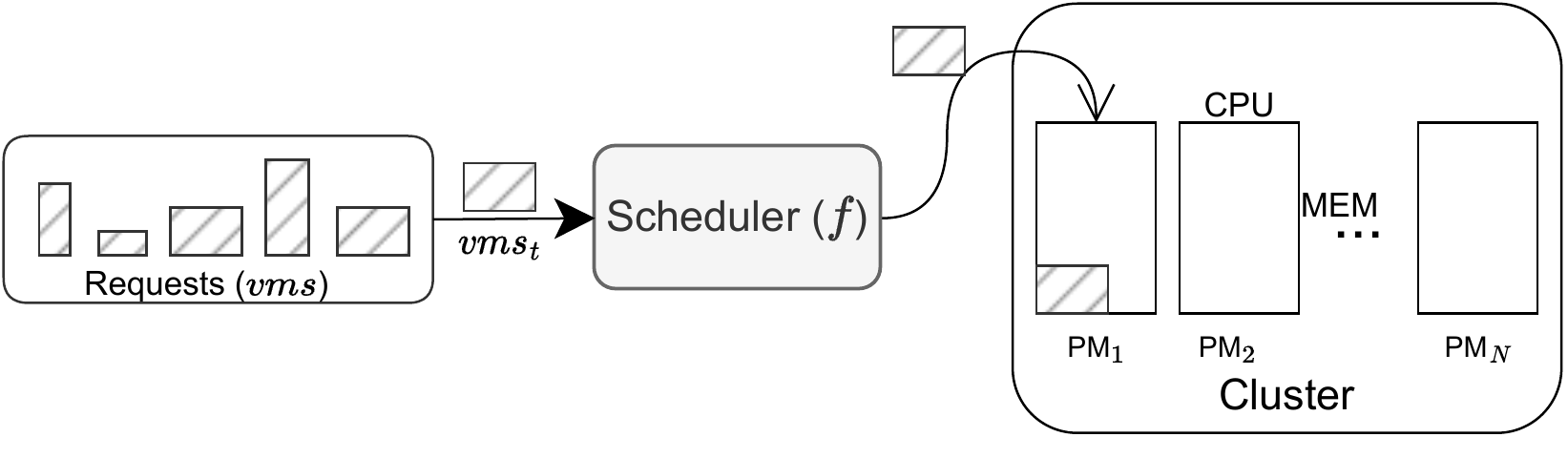}}
\caption{Scheduling Illustration. The scheduler keeps scheduling incoming virtual machine requests to one of the PMs in the cluster.}
\label{fig: sched}
\end{figure}
The scheduling problem in cloud computing, as shown in Fig.~\ref{fig: sched}, usually consists of three components: requests, scheduler, and cluster.
Consider a cluster that has $N$ PMs and each PM has $r^c$ CPU and $r^m$ memory resources.
A requests sequence consists of a sequence of virtual machine requests that need to be handled sequentially.
Each time the requests sequence delivers a request to the scheduler $f$, the scheduler selects which PM to handle the request in the cluster. 
When a PM handles a request, the PM will reserve resources for the request, and the using resources of the PM will increase accordingly.
If the using resources exceed the resource capacity, the scheduling process terminates, and no request will be handled after that. 
Denote the number of handled allocation requests before termination as scheduling length. 
The main goal of the scheduling problem is to find a scheduler $f$ that can handle as many allocation requests as possible (maximize scheduling length) with the cluster. 
The notations for the problem formulation are summarized in Table~\ref{tb:notations}.

\begin{table}[htbp] 
  \centering
  \caption{Notations.}
    \begin{tabular}{ll}
    \toprule
    \textbf{Symbol} & \textbf{Definition} \\
    \midrule
    $N$ & The total number of PMs in the cluster \\
    $r^c, r^m$ & The capacity of CPU and memory resources on a PM \\
    $u^c_i, u^m_i$  & The using CPU and memory resources of PM $i$\\
    $f$ & The scheduling method for allocating the incoming requests \\
    \bottomrule
    \end{tabular}%
    \label{tb:notations}
\end{table}%

According to the actual deployment of Huawei Cloud~\citep{sheng2022learning}, the PMs adopt multi Non-Uniform Memory Access (NUMA) architecture which allows allocation based on the VM size: a small size VM needs to be placed on a single NUMA of a PM while a large one needs to be divided equally and created on the multiple NUMAs of a PM. 
It should be noted that this paper does not change the scheduler directly. 
We explain and describe our approach ignoring the multi-NUMA setting to ease the understanding but take experiments on the practical multi-NUMA settings.

\section{Heterogeneity Analysis}
\label{sec: heterogeneity}
\begin{figure}[htbp]
\centerline{\includegraphics[width=.48\textwidth]{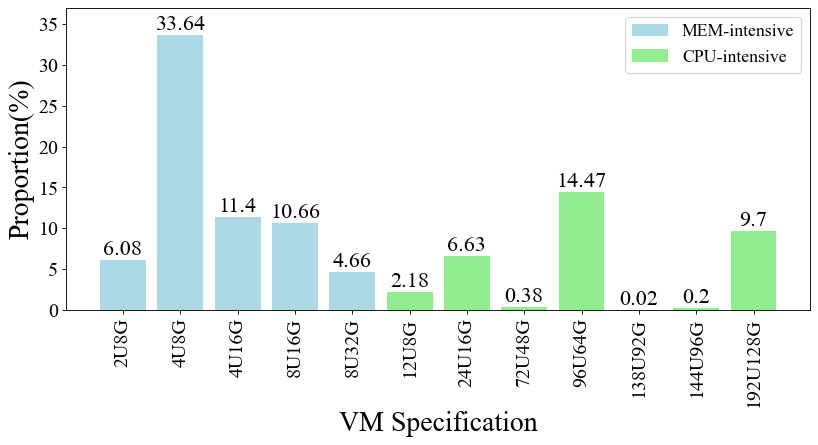}}
\caption{Statistics of the requests in Huawei-East-1 dataset.}
\label{fig: quantities}
\end{figure}

This section takes the real traces collected from Huawei Cloud to analyze the heterogeneity impacts on scheduling performance. 
Fig.~\ref{fig: quantities} makes statistics on the specification of allocation requests and their proportion in the dataset. 
There are 12 types of VM specification for the requests, each VM specification has different requirements for the CPU and memory resources and occupies different proportions.
Some requests consume CPU even larger than 96 Cores, while others only require 2 Cores. 
There are three kinds of CPU-to-Memory ratios for all requests, \ie, 1:4, 1:2, and 3:2.
The CPU requirements and CPU-to-Memory ratios can be taken as the heterogeneity of the requests. 
The next question is how much these heterogeneities influence the scheduling performance. 

This paper configures all the PMs with 128U160G resources, considering the real cloud setting and making the blurring.
To sufficiently evaluate the scheduling performance, the cluster is determined with 100 PMs, and $60$ start points are uniformly sampled from the dataset.
The scheduling performance of a scheduler can then be measured by calculating the average and quartiles of its allocation performances (scheduling lengths) in these $60$ start points with the cluster. 
In Table~\ref{tb: heter}, the ``all" filter rows show different schedulers' allocation performances respectively (the average, first quartile, median, and third quartile of allocation length). 
It should be noted that the optimal scheduler here is approximated by assembling offline random searches and other existing schedulers. 
It can be concluded that the existing heuristic schedulers have a significant potential for improvement.

Then to evaluate the impacts of the heterogeneity, two heterogeneities are summarized: the requirement of CPU and the resource preference. 
For the requirement of CPU, a request is categorized as a large CPU request if it requires more than $32$U practically.
Others are then called small requests. 
For the resource preference, if the CPU-to-Memory ratio of a request is larger than the CPU-to-Memory ratio of the PM setting, the request is denoted as CPU-Intensive one while others are denoted as MEM-Intensive requests as colored in Fig.~\ref{fig: quantities}. 

To evaluate the impact of the two heterogeneities, 
a \textit{filter-and-examination} way is proposed. 
It filters out one heterogeneity from the dataset by dividing the dataset into two homogeneous datasets. 
Suppose heuristic schedulers get close to the optimal scheduler under the two homogeneous datasets. In that case, the heterogeneity is called the impactful one.
As shown in Table~\ref{tb: heter}, the heuristic scheduler still has a large gap with the optimal scheduler when it filters out the CPU requirement heterogeneity. 
Thus, the CPU requirement is not an impactful heterogeneity. 
When filtering the dataset with only CPU-Intensive or MEM-Intensive requests, heuristic schedulers achieve similar performances as the optimal scheduler, as shown in Table~\ref{tb: heter}.
Thus resource preference can be taken as an impactful heterogeneity.
If the heterogeneity is reduced, the heuristic scheduler can be intensified accordingly. 

\begin{table}[htbp]
  \centering
  \caption{Comparison by filtering out heterogeneity}
  \scalebox{0.85}{
    \begin{tabular}{|c|c|cccc|}
    \toprule
    \multirow{2}[4]{*}{\textbf{Filter}} & \multirow{2}[4]{*}{\textbf{Algorithm}} & \multicolumn{4}{c|}{\textbf{Length}} \\
\cmidrule{3-6}      &   & \textbf{Average} & \textbf{Quartile 1} & \textbf{Quartile 2} & \textbf{Quartile 3} \\
    \midrule
    \multirow{5}[4]{*}{All} & Optimal & $\textbf{3093.38}$ & $\textbf{2002.25}$ & $\textbf{3028.00}$ & $\textbf{4172.75}$ \\
      & IB & $2855.17$ & $1997.75$ & $2668.00$ & $3810.50$ \\
      & BF & $2773.65$ & $1894.25$ & $2500.50$ & $3771.50$ \\
      & FF & $2512.23$ & $1537.25$ & $2381.00$ & $3389.75$ \\
      & BF2 & $2716.33$ & $1896.00$ & $2473.50$ & $3747.75$ \\
    \midrule
    \multirow{5}[2]{*}{CPU-Intensive} & Optimal & $\textbf{1154.33}$ & $\textbf{749.22}$ & $\textbf{1213.50}$ & $\textbf{1522.22}$ \\
      & IB & $1154.21$ & $\textbf{749.22}$ & $\textbf{1213.50}$ & $\textbf{1522.22}$ \\
      & BF & $1104.90$ & $747.75$ & $1082.00$ & $1494.23$ \\
      & FF & $1041.90$ & $724.00$ & $1050.50$ & $1302.21$ \\
      & BF2 & $1104.90$ & $747.75$ & $1082.00$ & $1494.23$ \\
    \midrule
    \multirow{5}[2]{*}{MEM-Intensive} & Optimal & $\textbf{5866.23}$ & $\textbf{5098.20}$ & $\textbf{5454.00}$ & $\textbf{7104.20}$ \\
      & IB & $5864.75$ & $\textbf{5098.20}$ & $5453.50$ & $\textbf{7104.20}$ \\
      & BF & $5866.00$ & $\textbf{5098.20}$ & $5453.50$ & $\textbf{7104.20}$ \\
      & FF & $5865.86$ & $\textbf{5098.20}$ & $\textbf{5454.00}$ & $\textbf{7104.20}$ \\
      & BF2 & $5866.10$ & $\textbf{5098.20}$ & $\textbf{5454.00}$ & $\textbf{7104.20}$ \\
    \midrule
    \multirow{5}[2]{*}{Small} & Optimal & $\textbf{4800.85}$ & $\textbf{4300.22}$ & $\textbf{4780.00}$ & $\textbf{5424.50}$ \\
      & IB & $4652.23$ & $4268.50$ & $4706.00$ & $5164.50$ \\
      & BF & $4361.2$5 & $3590.75$ & $4434.50$ & $4870.00$ \\
      & FF & $4367.50$ & $3595.50$ & $4429.00$ & $4879.68$ \\
      & BF2 & $4367.13$ & $3589.00$ & $4426.00$ & $4875.68$ \\
    \midrule
    \multirow{5}[2]{*}{Large} & Optimal & \textbf{901.60} & \textbf{486.00} & \textbf{894.00} & \textbf{1252.00} \\
      & IB & \textbf{901.60} & \textbf{486.00} & \textbf{894.00} & \textbf{1252.00} \\
      & BF & \textbf{901.60} & \textbf{486.00} & \textbf{894.00} & \textbf{1252.00} \\
      & FF & \textbf{901.60} & \textbf{486.00} & \textbf{894.00} & \textbf{1252.00} \\
      & BF2 & \textbf{901.60} & \textbf{486.00} & \textbf{894.00} & \textbf{1252.00} \\
    \bottomrule
    \end{tabular}%
    }
    \label{tb: heter}
\end{table}%

\begin{figure}[htbp]
\subfigure[MEM-Intensive.]{
\begin{minipage}[t]{0.31\linewidth}
\centering
\includegraphics[width=2.5cm]{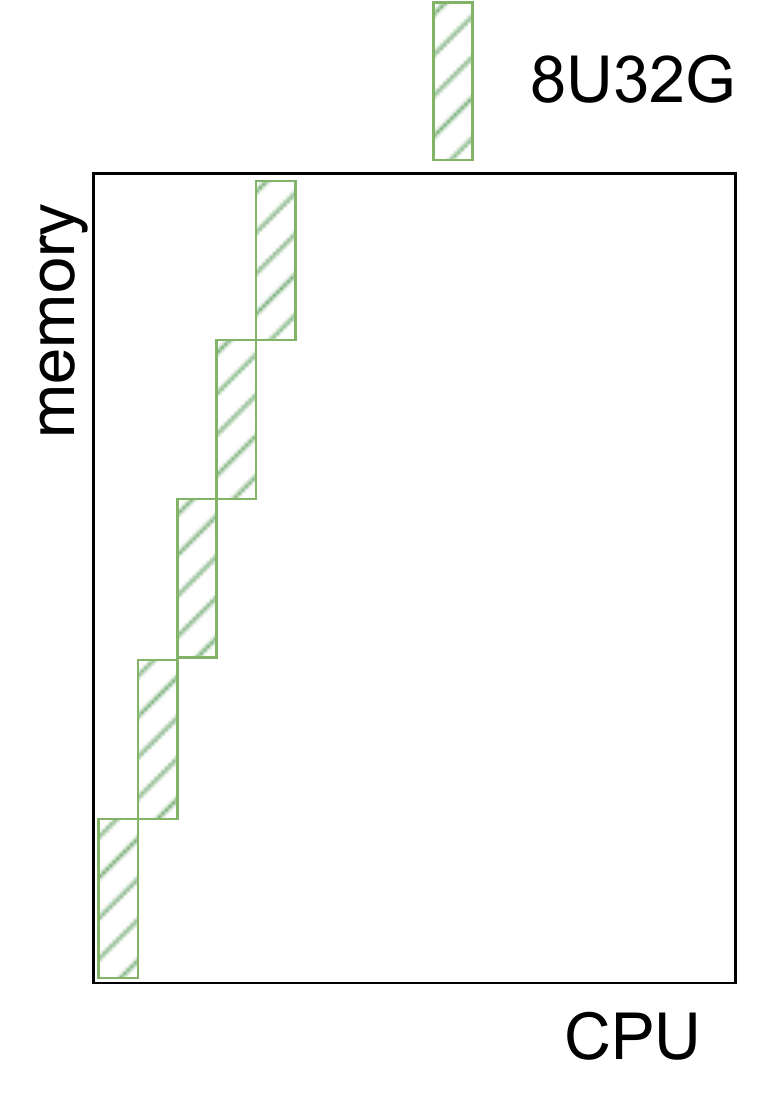}
\label{fig: uneven-a}
\end{minipage}%
}
\subfigure[CPU-Intensive.]{
\begin{minipage}[t]{0.31\linewidth}
\centering
\includegraphics[width=2.47cm]{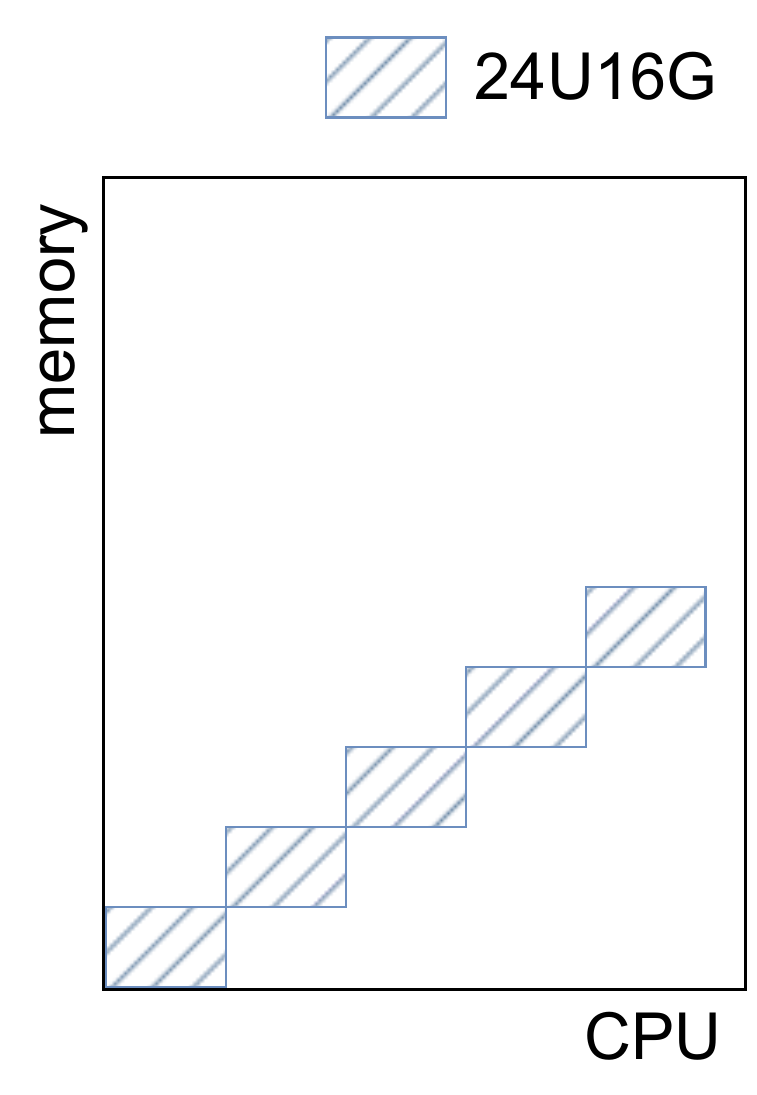}
\label{fig: uneven-b}
\end{minipage}%
}
\subfigure[Balanced.]{
\begin{minipage}[t]{0.31\linewidth}
\centering
\includegraphics[width=2.54cm]{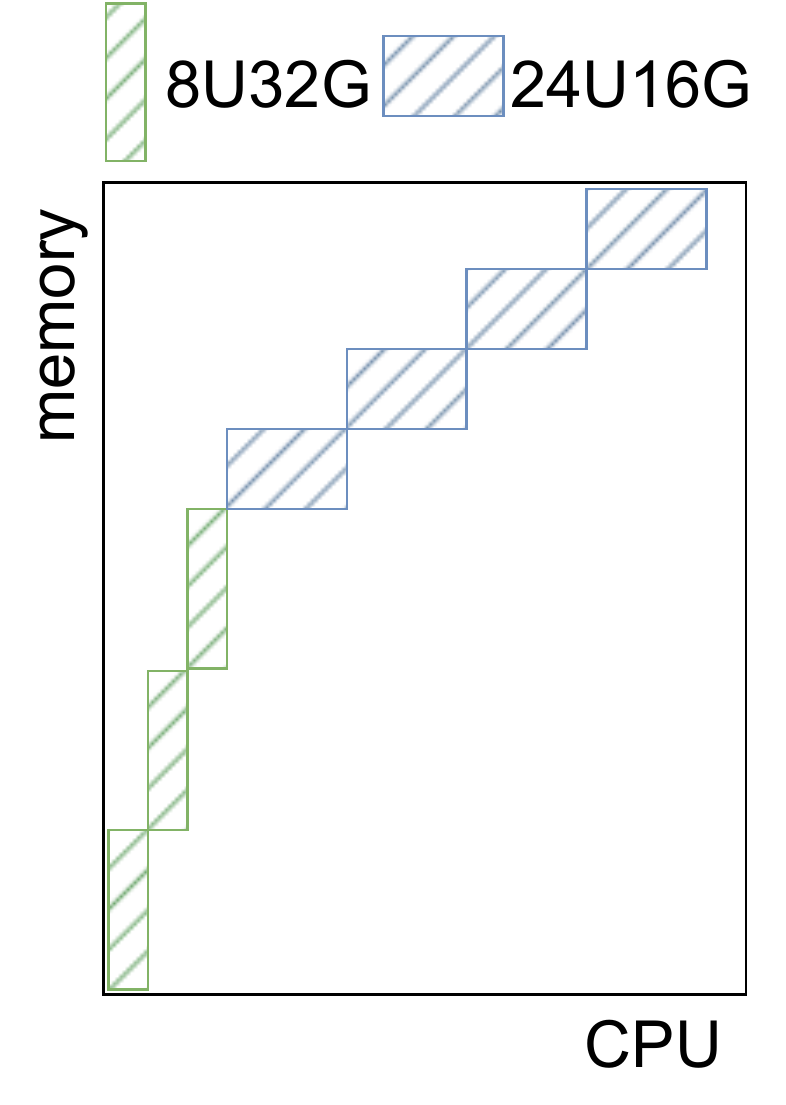}
\label{fig: uneven-c}
\end{minipage}%
}
\caption{Cases for uneven usage of resources. (a) corresponds to the waste of CPU resources caused by placing only MEM-Intensive VMs, and (b) corresponds to the waste of memory resources caused by placing only CPU-Intensive VMs. (c) corresponds to the result of the placement of two role VMs evenly.}
\label{fig: uneven}
\end{figure}

We visualize VM scheduling to further understand how resource preference heterogeneity contributes to performance downgrading.
Take the First-Fit scheduler as an example, its behavior is mainly decided by the order of incoming requests, and many unexpected situations would happen.
Three typical PM statuses are visualized in Fig.~\ref{fig: uneven}. 
In Fig.~\ref{fig: uneven-a}, only MEM-Intensive requests are allocated, and many CPU resources (88U) are wasted due to no memory supported to handle more requests.
In Fig.~\ref{fig: uneven-b}, memory resources are mainly wasted (96G) due to only allocating CPU-Intensive requests.
Fig.~\ref{fig: uneven-c} shows an ideal situation where both CPU-Intensive and MEM-Intensive requests are placed in a single PM, and resources are better utilized.

This paper introduces a novel metric: at least waste (ALW) to better characterize the waste on CPU and memory resources.
Denote the VM specification set as $\mathcal{V}$, and each VM specification $v\in \mathcal{V}$ requires $v^c$ CPU and $v^m$ memory resources.
The ALW on CPU resources and memory resources on PM $j$ with $u_j^c$ CPU and $u_j^m$ memory resources are defined as: 
\begin{equation}
\begin{aligned}
ALW^c(\delta^c, \delta^m, \mathcal{V}) &= \min_{v \in \mathcal{V}} \left\{ \delta^c - v^c \min(\lfloor \frac{\delta^c}{v^c} \rfloor, \lfloor \frac{\delta^m}{v^m} \rfloor) \right\}, \\
ALW^m(\delta^c, \delta^m, \mathcal{V}) &= \min_{v \in \mathcal{V}} \left\{ \delta^m - v^m \min(\lfloor \frac{\delta^m}{v^m} \rfloor, \lfloor \frac{\delta^c}{v^c} \rfloor) \right\},
\end{aligned}
\label{eq: ALW}
\end{equation}
where $\delta^c = r^c_j-u^c_j$ and $\delta^m = r^m_j - u^m_j$.
The ALW on CPU resources \eqref{eq: ALW} aims to calculate the at least waste of CPU given the remaining resources. 
The ALW on the CPU of the whole cluster is the sum of all the ALW of each PM in the cluster. 
Table~\ref{tb: waste} statics the ALW for different schedulers with the `all` filter. 
The second column shows the average allocation length. The third and fourth columns show the average and standard deviation of ALW on memory and CPU resources of the whole cluster.
There exists a large potential for existing schedulers to reduce ALW wastes.

\begin{table}[htbp]
  \centering
  \caption{ALW statics of different baselines. }
    \begin{tabular}{cccc}
    \toprule
    \textbf{Algorithm} & \textbf{Length} & \textbf{ALW on Memory} & \textbf{ALW on CPU} \\
    \midrule
    Optimal & $\textbf{3093.38}$ & $\textbf{8.39($\pm$7.31)}$ & $\textbf{1.28($\pm$2.38)}$ \\
    IB & $2855.17$ & $23.32(\pm14.37)$ & $5.41(\pm5.94)$ \\
    BF & $2773.65$ & $17.79(\pm10.70)$ & $6.36(\pm5.57)$ \\
    FF & $2512.23$ & $14.62(\pm12.06)$ & $10.29(\pm5.77)$ \\
    BF2 & $2716.33$ & $18.39(\pm11.00)$ & $6.98(\pm6.16)$ \\
    \bottomrule
    \end{tabular}%
    \label{tb: waste}
\end{table}%

\section{{\sc{ReAssigner}}: Resource Assigner}
\label{sec: approach}
\subsection{The Main Framework}
\begin{figure*}[htbp]
\centerline{\includegraphics[width=.85\textwidth]{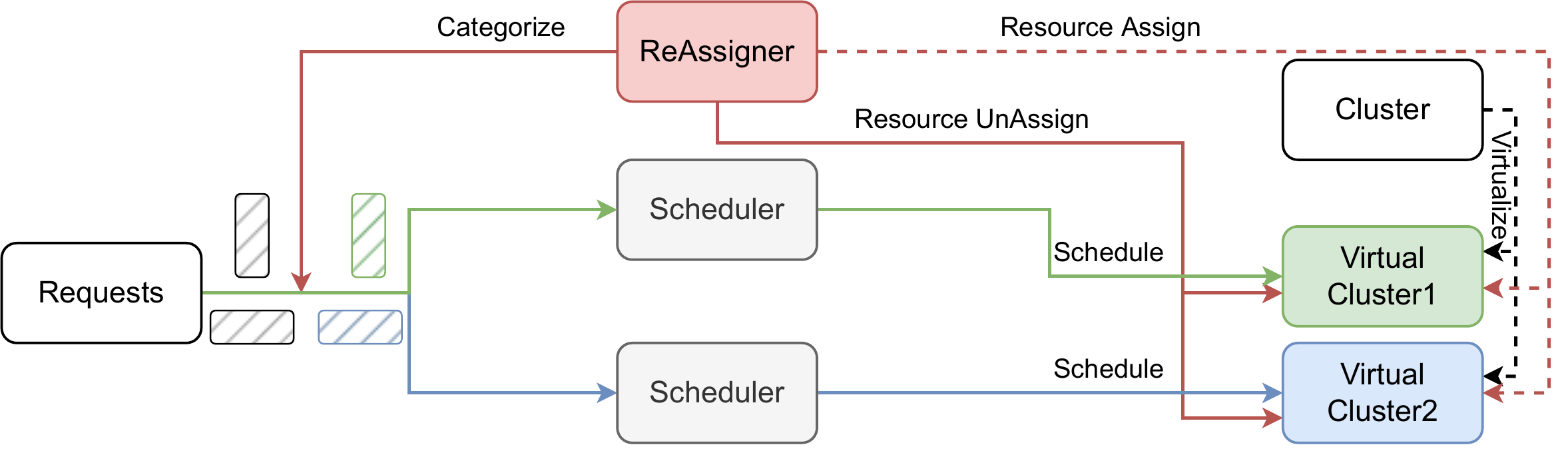}}
\caption{Flow diagram for {\sc{ReAssigner}}. The dashed lines are only activated at start and the solid lines activate for each request. The green hatch and blue hatch are two meta-type requests. {\sc{ReAssigner}} first assigns two virtual clusters. When a request comes, {\sc{ReAssigner}} categorizes it based on the meta-type and makes unassign decision on the virtual clusters. Then scheduler schedules the request to its corresponding virtual cluster.}
\label{fig: framework}
\end{figure*}

This section describes the overall framework of our Resource Assigner ({\sc{ReAssigner}}). 
The main intuition behind the {\sc{ReAssigner}} is that the impacts of heterogeneity in requests on scheduling can be reduced by pre-assigning roles to resources and letting resources of the same role form a virtual cluster to handle homogeneous requests. 
As shown in Fig.~\ref{fig: framework}, {\sc{ReAssigner}} consists three sub-modules: {\em{assign}}, {\em{unassign}}, and {\em{categorize}}.
{\sc{ReAssigner}} first establishes two virtual  clusters and {\em{assigns}} roles (CPU-Intensive or MEM-Intensive) to the resources in the virtual clusters. 
Each time a request comes, {\sc{ReAssigner}} categorizes it as what role (meta-type\footnote{The meta-type of a request is its role in this paper.}) it belongs to and lets the scheduler schedule it on the corresponding virtual cluster.
Although assigning roles to resources reduces the impact of requests heterogeneity, it restricts the scheduling space\footnote{The scheduling space here denotes how many positions the scheduler can schedule the requests on.} which limits the scheduling performance. 
{\sc{ReAssigner}} adaptively unassigns the roles on the resources to intensify the scheduling performance further.

\subsection{Categorization}
This section presents how {\sc{ReAssigner}} categories virtual machine requests.
As shown in Section~\ref{sec: heterogeneity}, there are many specifications of requests.
Intuitively, requests with the same CPU-to-Memory ratio can be clustered into the same types.
Directly categorizing the requests to their flavor types can result in high complexity on the later resource assignment. 
As analyzed in Section~\ref{sec: heterogeneity}, the mixture of different preferences critically impacts scheduling performance.
{\sc{ReAssigner}} proposes a coarse meta-type choosing: CPU-Intensive type and MEM-Intensive type. 
A request ($v\in \mathcal{V}$) is categorized as CPU-Intensive if its CPU-to-Memory ratio is larger than CPU-to-Memory ratio of the PM setting ($\frac{v^c}{v^m}\geq \frac{r^c}{r^m}$).
Others are then categorized as MEM-Intensive. 
The CPU-Intensive and MEM-Intensive requests sets are denoted as $\mathcal{V}(ci)$ and $\mathcal{V}(mi)$, respectively.

\subsection{Virtual Cluster and Assign Scheme}
This section describes how {\em{assign}} works, considering the cluster has $N$ PMs, each PM has $r^c$ CPU and $r^m$ memory.
Two virtual clusters (CPU-Intensive and MEM-Intensive Virtual Clusters) are established, and each cluster has $N$ VPMs.
Each VPM is initiated with $r^c$ CPU and $r^m$ memory, and their roles are not assigned. 
The resources with no roles specified are shared among virtual clusters and can be used to handle all the virtual machines in $\mathcal{V}$.
When the resources without roles are used by one virtual cluster, the other resources of the virtual cluster are synchronously used. 
For CPU-Intensive virtual cluster, it can assign a CPU-Intensive role to its resources and claim assigned resources only be allocate-able for CPU-Intensive virtual machines ($\mathcal{V}(ci)$). 
For MEM-Intensive clusters, it assigns MEM-Intensive roles to resources and claims to sit for MEM-Intensive virtual machines ($\mathcal{V}(mi)$). 

After defining the virtual clusters, a critical question follows: how to assign roles to resources?
{\sc{ReAssigner}} starts from three heuristics: 1) All PMs have the same resources assignment at initialization, 2) the resources assignment should minimize the ALW on both roles, and 3) the resources with a role in a VPM should at least be able to handle one largest VM request of role.
The {\em{assign}} sub-module is to find the number of CPU ($c^1$), memory ($m^1$) that should be assigned as CPU-Intensive and the number of CPU ($c^2$) and memory ($m^2$) that should be assigned as MEM-Intensive for each PM that can reduce the ALW wastes. 
The assignment problem can be formulated as follows:

\begin{equation}
\begin{small} 
\begin{aligned}
\min_{c^1, m^1, c^2, m^2} &\lambda \left[ ALW^c \left( c^1, m^1, \mathcal{V}(ci) \right) + ALW^c \left[ c^2, m^2, \mathcal{V}(mi) \right) \right]\\
+ (1-\lambda) &\left[ ALW^m \left( c^1, m^1, \mathcal{V}(ci) \right) + ALW^m \left( c^2, m^2, \mathcal{V}(mi) \right) \right] \\
\text{s.t.} \quad & 0 \leq c^1, c^2 \leq r^c, \quad 0 \leq m^1, m^2 \leq r^m, \\
& c^1 + c^2 = r^c, \quad m^1 + m^2 = r^m, \ \\
& 0 \leq \lambda \leq 1. 
\end{aligned}
\label{set_masking}
\end{small}
\end{equation}

where $\lambda$ is a weight factor that trades off between memory waste and CPU waste.
This problem can be solved by {\em{Combinatorial Optimization}} solvers, \eg, Gurobi~\citep{gurobi2018gurobi}.

\begin{figure}[htbp]
\centerline{\includegraphics[width=.48\textwidth]{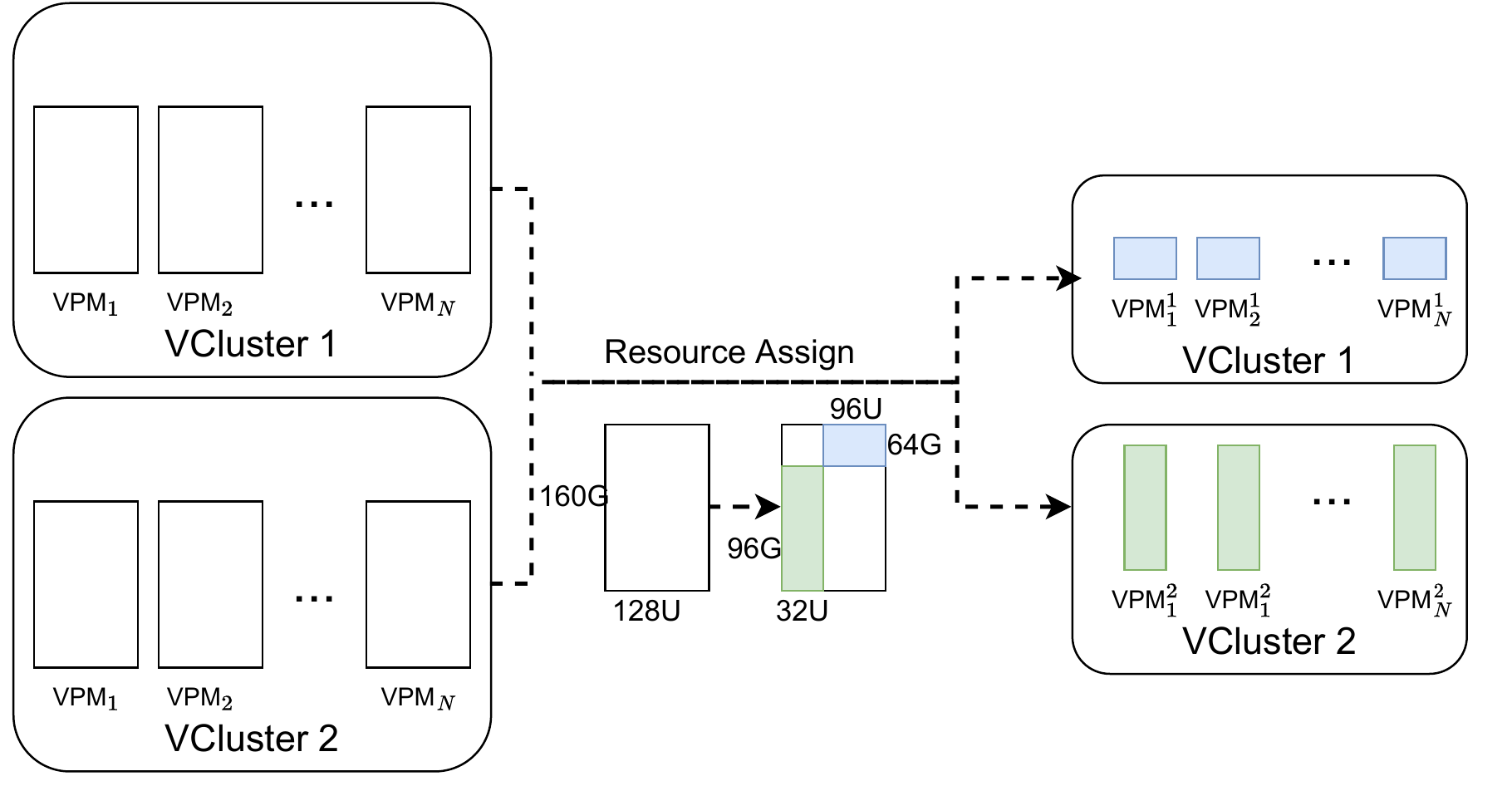}}
\caption{Resource Assign Illustration. The rectangles are the physical machines. Different solid color in the PM represents the different meta-types and the white color is the flexible resources. This figure shows the process of assigning $96$U$64$G resources to VCluster~$1$ and the others to VCluster~$2$ for each physical machine.}
\label{fig: assign}
\end{figure}

Generally, $\lambda$ needs to be specified based on the domain requirements. 
In our specific setting, considering the real cloud setting and making the blurring, each PM has $128$U$160$G resources, and the VM specification set is listed in the horizontal axis of Fig.~\ref{fig: quantities}.
There is an optimal solution ($c^1=96, c^2=32, m^1=64, m^2=96$) for arbitrary $\lambda$. 
This can be easily verified due to the objective in \eqref{set_masking} being $0$ with the aforementioned solution.
Thus {\sc{ReAssigner}} assigns the CPU-Intensive role to 96U64G resources and the MEM-Intensive role to the remaining 32U96G resources for each PM.
As shown in Fig.~\ref{fig: assign}, VCluster~1 assigns the CPU-Intensive role with blue hatches, and VCluster~2 assigns the MEM-Intensive role with green hatches.

\subsection{Unassign Scheme}

\begin{figure}[htbp]
\centerline{\includegraphics[width=.48\textwidth]{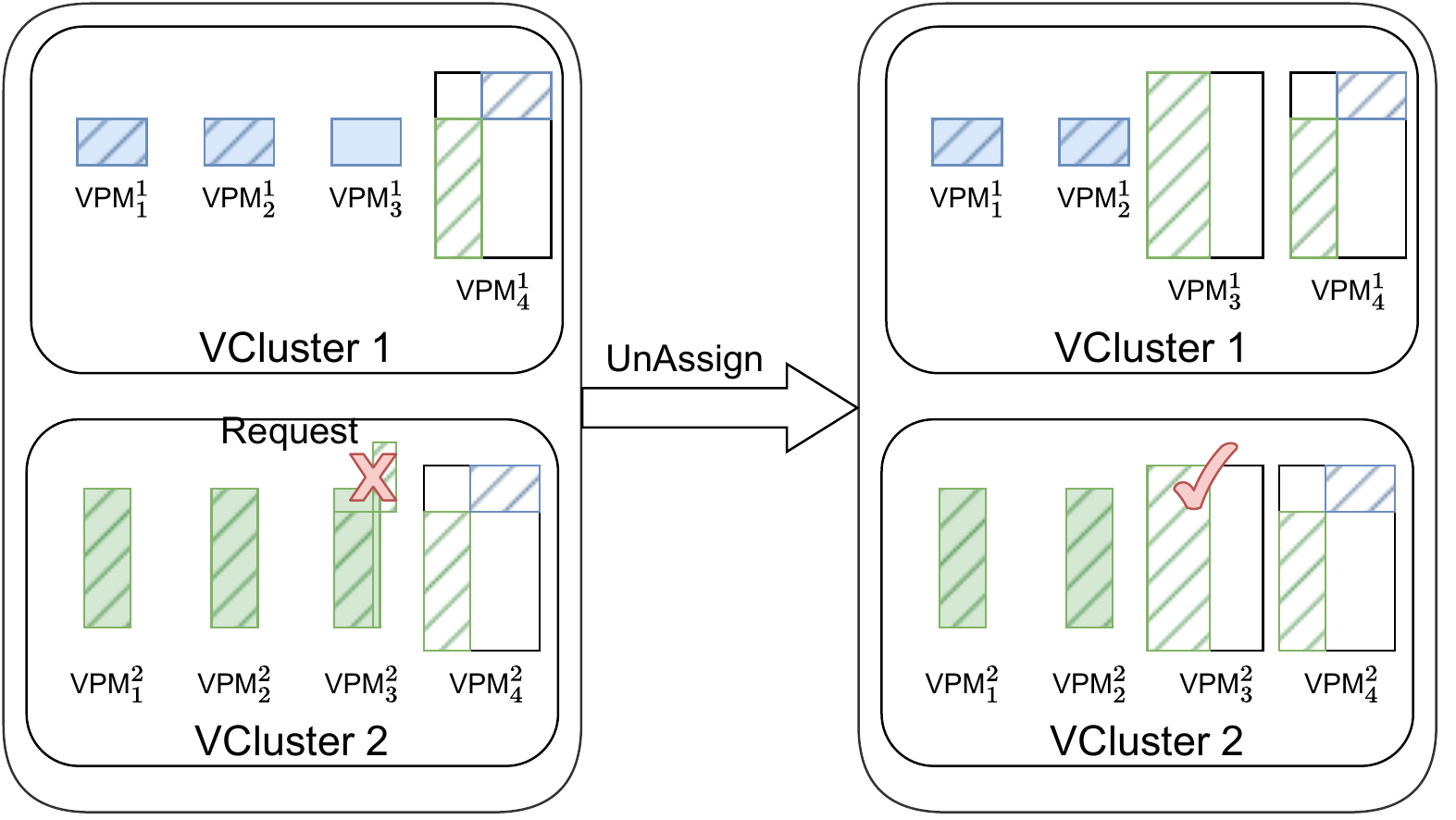}}
\caption{Unassignment at emergent situation. The left side shows that the VCluster~$2$ can not handle the current request with meta-type~$2$, which we call it the emergent situation. The {\em{unassign}} module unassigns the VPM$_3^1$ and VPM$_3^2$. The current request gets handled.}
\label{fig: unassign2}
\end{figure}

Although the assignment scheme helps the scheduler better utilize the PM's resources, it restricts the scheduling space of the scheduler and can thus degrade the scheduling performance. 
As shown in Fig.~\ref{fig: unassign2}, VCluster~$2$ does not have enough resources to place the incoming MEM-Intensive request. 
However, when we scan the physical servers, the PM$_3$ can allocate the request. 
Such a scheduling performance degrader is mainly caused by the smaller scheduling space defined in the assignment scheme. 
This section seeks a scheme that can enjoy the benefits of balanced allocation and reduce the effect of limited scheduling space: unassign. 
The unassign selects a PM and unassigns the two VPMs virtualized from it. 
When both the VPMs get unassigned, the two VPMs share all the resources from the PM, which means both virtual clusters can utilize the PM resources. 
This enables us to reduce the limitation of assignments. 
It also should be noted that if we unassign all the VPMs, our {\sc{ReAssigner}} degrades to the originally existing scheduler.
The next question is how to trade off the assignment and unassignment. 

\begin{figure}[htbp]
\centerline{\includegraphics[width=.48\textwidth]{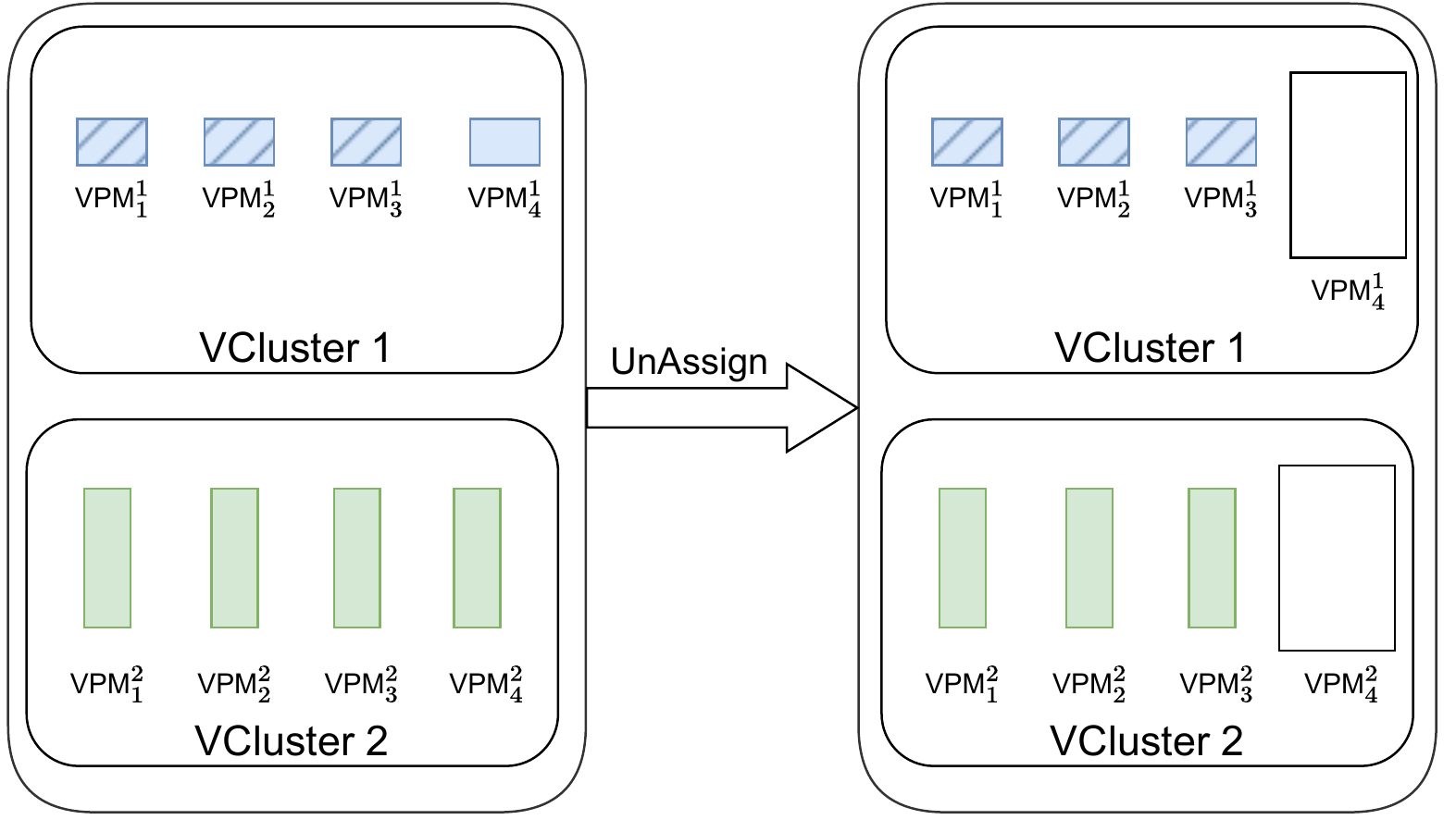}}
\caption{Unassignment at Unbalanced utilization. The hatch is the using resources. The VCluster~$1$ has utilized $75\%$ resources while VCluster~$2$ has used $0$ resources, which is highly unbalanced between virtual clusters. The {\em{unassign}} module then unassigns the VPM$_4^1$ and VPM$_4^2$ to the shareable resources.}
\label{fig: unassign}
\end{figure}

This paper studies two conditions that {\sc{ReAssigner}} needs to make unassignment: 1) Emergent situation on allocation, and 2) Unbalanced utilization between virtual clusters. For the first condition, if the resources of VPMs corresponding to the next VM are not enough, {\sc{ReAssigner}} will unassign the first PM that can place the VM after the unassignment.
For the second condition, when one virtual cluster has utilized a large proportion of resources, and the other one nearly consumes no resources, it indicates one of the meta-type requests dominates. 
More resources should be left to the former one. 
{\sc{ReAssigner}} will unassign one of the PMs, whose two VPMs are all empty, to be shareable for two meta-type VMs.
As shown in Fig.~\ref{fig: unassign}, three of the VPMs in the VCluster~$1$ get fully utilized while all the VPMs in the VCluster~$2$ are idle. 
It is reasonable to unassign the PM$_4$ to share resources to handle CPU-Intensive requests. 

Denote the using CPU and memory for the PM $j$ in CPU-Intensive VCluster as $u_j^c(ci), u_j^m(ci)$, the using CPU and memory for MEM-Intensive VCluster as $u_j^c(mi), u_j^m(mi)$, and the number of PMs unassigned with unbalanced utilization as $N_{unassign}$. 
The usage of the PMs on the VCluster can be roughly calculated by dividing the total usage of CPU or memory by the assigned resources.
Then the difference of usage of the PMs in two VClusters can be roughly subtracted. 
The PMs are unassigned in advance to deal with the imbalances that have arrived, so the imbalance ($\Gamma_{unbalance}$) can be calculated by subtracting the minimum of differences based on CPU and memory by the number of unassigned PMs, as shown in (\ref{imbalance of two meta types}). 

\begin{equation} \label{imbalance of two meta types}
\begin{gathered}
\Gamma_{imbalance} = \min \left\{ \bigg | \frac{\sum_{j} u_j^c(ci)}{c^1} - \frac{\sum_{j} u_j^c(mi)}{c^2} \bigg |,\right.\\
\qquad \left. \bigg |\frac{\sum_{j} u_j^m(ci)}{m^1} - \frac{\sum_{j} u_j^m(mi)}{m^2} \bigg | \right\} - N_{unassign}.
\end{gathered}
\end{equation}
If the imbalance exceeds an imbalance threshold $\alpha$, {\sc{ReAssigner}} will unassign two VPMs that belong to the same PM as shown in Fig.~\ref{fig: unassign2}.

The overall process of the proposed {\sc{ReAssign}} scheme is shown in Algorithm.~\ref{al: reassigner}.

\begin{figure}[!t]
    \begin{algorithm}[H]
        \caption{{\sc{ReAssigner}}}
        \begin{algorithmic}[1]
            \REQUIRE The total number of PMs $N$. The capacity of CPU and memory on a PM $r^c, r^m$. The handled length $L=0$        
            \STATE {Define meta-type sets, $\mathcal{V}(ci)$ for CPU-Intensive and $\mathcal{V}(mi)$ for MEM-Intensive, and calculate CPU-to-Memory ratios $r^1, r^2$ of $\mathcal{V}(ci)$ and $\mathcal{V}(mi)$.}
            \STATE {Set $c^1, c^2, m^1, m^2$ by (\ref{set_masking}), and unassign the cluster to VCluster~$1$ and VCluster~$2$.}
            \WHILE {The cluster can place the incoming $v m$}
                \IF{$v m$ is a deletion request}
        		\STATE {Release the resources.}
        		\ENDIF
                \IF{$v m$ is a creation request}
        		\STATE {Choose the PM $i$ according to its meta-type based on the scheduler.}
                \STATE {$u_{i}^c, u_{i}^m = f\left(u_{i}^c, u_{i}^m, v m\right)$}
                \STATE {$L=L+1$}
                \IF{Resources are not enough for the next VM}
                \STATE {$Unassign$ the first PM, which can place the VM after the unassignment.}
                \ENDIF
        		\IF{$\Gamma_{imbalance} \geq \alpha$}
                \STATE {$Unassign$ one of the PMs, whose two VPMs are all empty.}
        		\ENDIF
        		\ENDIF
            \ENDWHILE
            \RETURN $L$, $ALW^c$ and $ALW^m$
        \end{algorithmic}
        \label{al: reassigner}
    \end{algorithm}
\end{figure}

\section{Experiments}
\label{sec: evaluation}
\subsection{Environment settings and details of {\sc{ReAssigner}}}
This paper takes, Huawei-East-1\footnote{https://github.com/mail-ecnu/VMAgent}, the real dataset open-sourced by Huawei, as the main playground.
It contains one-month interaction data in one cluster of Huawei Cloud. 
To make a fair and sufficient comparison, $60$ start points are uniformly sampled from the dataset to construct rich testing scenarios.
For cluster configurations, this paper considers homogeneous PMs with $128$U$160$G resources, considering the real cloud setting and making certain blurring.

\begin{table}[htbp]
  \centering
  \caption{Resources Statics of two meta-type Requests.}
  \scalebox{0.9}{
    \begin{tabular}{cccc}
    \toprule
    \textbf{Meta-type} & \textbf{CPU-to-Memory} & \textbf{Total CPU(U)} & \textbf{Total memory(G)} \\
    \midrule
    CPU-Intensive & 3:2 & 2217762 & 1478508 \\
    MEM-Intensive & 1:2,1:4 & 183000 & 470456 \\
    \bottomrule
    \end{tabular}%
    }
  \label{tab:total resource}%
\end{table}%

Resources statics of two meta-type requests are shown in Table.\ref{tab:total resource}. After solving the optimization problem (\ref{set_masking}), 96U64G resources are assigned for the CPU-Intensive role and 32U96G for the MEM-Intensive role.
The threshold $\alpha$ for unassignment is a hyperparameter, and it is selected through grid searches.
By searching $\alpha$ in $\left[ {\rm{0.05N, 0.1N, 0.2N, 0.3N}} \right]$ where $N$ is the number of PMs, $0.3N$ achieves the best performance on the length and is used in the rest of the experiment.
\begin{table*}[htbp]
  \centering
  \caption{Performance comparison under different cluster sizes}
    \begin{tabular}{|c|c|cccc|cc|}
    \toprule
    \multirow{2}[4]{*}{\textbf{Number of PMs}} & \multirow{2}[4]{*}{\textbf{Algorithm}} & \multicolumn{4}{c|}{\textbf{Length}} & \multicolumn{2}{c|}{\textbf{ALW}} \\
\cmidrule{3-8}      &   & \textbf{Average} & \textbf{Quartile 1} & \textbf{Quartile 2} & \textbf{Quartile 3} & \textbf{Memory} & \textbf{CPU} \\
    \midrule
    \multirow{9}[10]{*}{50 PMs} & Optimal & \textit{$\textbf{1121.38}$} & \textit{$\textbf{478.00}$} & \textit{$\textbf{899.50}$} & \textit{$\textbf{1748.75}$} & \textit{$\textbf{6.22($\pm$7.37)}$} & \textit{$\textbf{0.72($\pm$2.30)}$} \\
\cmidrule{2-8}      & IB & $\textbf{1077.70}$ & $\textbf{478.00}$ & $\textbf{899.50}$ & $\textbf{1640.50}$ & $18.97(\pm17.52)$ & $3.06(\pm4.29)$ \\
      & IB+RA & $1075.20$ & $449.50$ & $852.50$ & $1624.25$ & $\textbf{8.59($\pm$9.70)}$ & $\textbf{0.74($\pm$2.31)}$ \\
\cmidrule{2-8}      & BF & $1023.83$ & $\textbf{476.75}$ & $\textbf{836.50}$ & $\textbf{1475.75}$ & $14.91(\pm12.31)$ & $4.28(\pm4.96)$ \\
      & BF+RA & $\textbf{1040.88}$ & $448.75$ & $834.00$ & $1401.50$ & $\textbf{9.89($\pm$7.22)}$ & $\textbf{1.94($\pm$3.50)}$ \\
\cmidrule{2-8}      & FF & $921.25$ & $\textbf{470.75}$ & $783.00$ & $1288.75$ & $10.85(\pm10.74)$ & $7.85(\pm5.84)$ \\
      & FF+RA & $\textbf{1038.30}$ & $448.75$ & $\textbf{834.00}$ & $\textbf{1400.00}$ & $\textbf{9.46($\pm$6.89)}$ & $\textbf{2.21($\pm$3.76)}$ \\
\cmidrule{2-8}      & BF2 & $1009.53$ & $\textbf{476.75}$ & $\textbf{836.00}$ & $1475.50$ & $14.51(\pm11.69)$ & $4.77(\pm5.55)$ \\
      & BF2+RA & $\textbf{1051.87}$ & $448.75$ & $834.00$ & $\textbf{1506.50}$ & $\textbf{9.95($\pm$7.32)}$ & $\textbf{2.06($\pm$3.61)}$ \\
    \midrule
    \multirow{9}[9]{*}{100 PMs} & Optimal & \textit{$\textbf{3093.38}$} & \textit{$\textbf{2002.25}$} & \textit{$\textbf{3028.00}$} & \textit{$\textbf{4172.75}$} & \textit{$\textbf{8.39($\pm$7.31)}$} & \textit{$\textbf{1.28($\pm$2.38)}$} \\
\cmidrule{2-8}      & IB & $2855.17$ & $\textbf{1997.75}$ & $2668.00$ & $3810.50$ & $23.32(\pm14.37)$ & $5.41(\pm5.94)$ \\
      & IB+RA & $\textbf{3031.78}$ & $1967.00$ & $\textbf{2831.50}$ & $\textbf{4172.75}$ & $\textbf{11.00($\pm$8.93)}$ & $\textbf{1.62($\pm$2.75)}$ \\
\cmidrule{2-8}      & BF & $2773.65$ & $1894.25$ & $2500.50$ & $3771.50$ & $17.79(\pm10.70)$ & $6.36(\pm5.57)$ \\
      & BF+RA & $\textbf{2917.83}$ & $\textbf{1933.50}$ & $\textbf{2702.50}$ & $\textbf{4018.25}$ & $\textbf{12.90($\pm$6.48)}$ & $\textbf{3.18($\pm$3.82)}$ \\
\cmidrule{2-8}      & FF & $2512.23$ & $1537.25$ & $2381.00$ & $3389.75$ & $14.62(\pm12.06)$ & $10.29(\pm5.77)$ \\
      & FF+RA & $\textbf{2881.62}$ & $\textbf{1933.50}$ & $\textbf{2654.00}$ & $\textbf{4009.50}$ & $\textbf{12.57($\pm$6.28)}$ & $\textbf{3.59($\pm$4.13)}$ \\
\cmidrule{2-8}      & BF2 & $2716.33$ & $1896.00$ & $2473.50$ & $3747.75$ & $18.39(\pm11.00)$ & $6.98(\pm6.16)$ \\
      & BF2+RA & $\textbf{2906.18}$ & $\textbf{1933.50}$ & $\textbf{2702.50}$ & $\textbf{4018.25}$ & $\textbf{12.92($\pm$6.48)}$ & $\textbf{3.31($\pm$3.91)}$ \\
    \midrule
    \multirow{9}[9]{*}{150 PMs} & Optimal & \textit{$\textbf{5144.28}$} & \textit{$\textbf{3528.00}$} & \textit{$\textbf{5196.50}$} & \textit{$\textbf{7021.00}$} & \textit{$\textbf{8.78($\pm$6.34)}$} & \textit{$\textbf{1.87($\pm$3.04)}$} \\
\cmidrule{2-8}      & IB & $4790.42$ & $\textbf{3523.50}$ & $4998.00$ & $6038.75$ & $25.15(\pm12.38)$ & $6.51(\pm7.41)$ \\
      & IB+RA & $\textbf{5113.83}$ & $3352.75$ & $\textbf{5190.50}$ & $\textbf{7021.00}$ & $\textbf{12.02($\pm$8.10)}$ & $\textbf{1.88($\pm$3.03)}$ \\
\cmidrule{2-8}      & BF & $4776.68$ & $3349.75$ & $4779.50$ & $6149.50$ & $19.00(\pm9.52)$ & $7.20(\pm6.01)$ \\
      & BF+RA & $\textbf{4979.32}$ & $\textbf{3352.75}$ & $\textbf{4985.00}$ & $\textbf{6593.75}$ & $\textbf{13.98($\pm$5.60)}$ & $\textbf{3.63($\pm$4.01)}$ \\
\cmidrule{2-8}      & FF & $4300.05$ & $2987.25$ & $4195.00$ & $5731.50$ & $15.04(\pm11.13)$ & $11.55(\pm6.04)$ \\
      & FF+RA & $\textbf{4920.18}$ & $\textbf{3348.25}$ & $\textbf{4731.00}$ & $\textbf{6573.25}$ & $\textbf{13.39($\pm$5.63)}$ & $\textbf{4.21($\pm$4.01)}$ \\
\cmidrule{2-8}      & BF2 & $4713.90$ & $3351.50$ & $4779.00$ & $6088.25$ & $19.88(\pm9.95)$ & $8.05(\pm6.83)$ \\
      & BF2+RA & $\textbf{4973.55}$ & $\textbf{3352.75}$ & $\textbf{4965.00}$ & $\textbf{6593.50}$ & $\textbf{14.02($\pm$5.64)}$ & $\textbf{3.72($\pm$4.00)}$ \\
    \bottomrule
    \end{tabular}%
  \label{tab:comparison}%
\end{table*}%

\begin{table*}[htbp]
  \centering
  \caption{Ablation with two types of unassignment given 100 PMs based on IB}
    \scalebox{1.03}{\begin{tabular}{|c|cccc|cc|}
    \toprule
    \multirow{2}[4]{*}{\textbf{Scheduler}} & \multicolumn{4}{c|}{\textbf{Length}} & \multicolumn{2}{c|}{\textbf{ALW}} \\
\cmidrule{2-7}      & \textbf{Average} & \textbf{Quartile 1} & \textbf{Quartile 2} & \textbf{Quartile 3} & \textbf{Memory} & \textbf{CPU} \\
    \midrule
    Ours & $\textbf{3031.78}$ & $\textbf{1967.00}$ & $\textbf{2831.50}$ & $4172.75$ & $11.00(\pm8.93)$ & $1.62(\pm2.75)$ \\
    w/o UnassignEmergent & $3026.92$ & $\textbf{1967.00}$ & $\textbf{2831.50}$ & $4172.75$ & $10.83(\pm8.63)$ & $1.59(\pm2.72)$ \\
    w/o UnassignUnbalanced & $2956.92$ & $1918.50$ & $2650.50$ & $\textbf{4173.25}$ & $5.09(\pm5.81)$ & $1.20(\pm2.69)$ \\
    w/o unassign & $2842.65$ & $1918.50$ & $2587.50$ & $4057.00$ & $\textbf{3.57($\pm$3.86)}$ & $\textbf{0.05($\pm$0.05)}$ \\
    \bottomrule
    \end{tabular}%
  \label{tab:ablation}%
  }
\end{table*}%

\begin{table*}[htbp]
  \centering
  \caption{Comparison with different assigning given 100 PMs based on IB}
    \begin{tabular}{|c|cccc|cc|}
    \toprule
    \multirow{2}[4]{*}{\textbf{Assignment}} &  \multicolumn{4}{c|}{\textbf{Length}} & \multicolumn{2}{c|}{\textbf{ALW}} \\
\cmidrule{2-7}      & \textbf{Average} & \textbf{Quartile 1} & \textbf{Quartile 2} & \textbf{Quartile 3} & \textbf{Memory} & \textbf{CPU} \\
    \midrule
    96U64G(Ours) & $\textbf{3031.78}$ & $1967.00$ & $\textbf{2831.50}$ & $\textbf{4172.75}$ & $11.00(\pm8.93)$ & $1.62(\pm2.75)$ \\
    108U72G & $2813.17$ & $1934.00$ & $2653.00$ & $3930.75$ & $8.26(\pm6.66)$ & $3.42(\pm4.95)$ \\
    120U80G & $2733.23$ & $\textbf{1996.00}$ & $2474.00$ & $3605.25$ & $24.98(\pm16.24)$ & $9.39(\pm8.77)$ \\
    84U56G & $2400.57$ & $1431.00$ & $2186.00$ & $3219.75$ & $\textbf{4.17($\pm$8.06)}$ & $\textbf{0.72($\pm$1.17)}$ \\
    \bottomrule
    \end{tabular}%
  \label{tab:diff_mask}%
\end{table*}%
\subsection{Metrics and Baselines}
To measure the performance of schedulers, a critical metric is the number of requests a scheduler can handle (length). 
Due to the diverse performance among different scenarios, the mean, first quartile, median, and third quartile of the length are taken as the primary metrics. 
When two schedulers achieve similar lengths, the ALW on CPU and memory \eqref{eq: ALW} at the termination state are considered as an indirect metric to show the placement situation.
If a scheduler has a smaller ALW but a similar length to others, the scheduler will be taken as a better one due to its better termination status.
It should also be noted that the ALW does not contribute to comparison if the scheduling performances are significantly different.

For baselines, this paper considers four heuristic schedulers: 
\begin{itemize}
\item FF (First-Fit), that is, to place the VM request to the first available PM.

\item BF (Best-Fit), that is, to place the VM request to the best PM with the highest CPU utilization among all the available PMs.

\item BF2~\citep{hadary2020protean}, the popular scheduling method for multi-dimension resources used by Protean.
It places the VM request based on the weighted score of each PM.

\item IB (Internal Baseline), the scheduling method used internally by Huawei.

\end{itemize}

The {\sc{ReAssigner}} are evaluated on these different schedulers to show its wide applicability.

\subsection{Performance comparison under different cluster size}\label{Performance comparison with baselines}

This section evaluates {\sc{ReAssigner}} under the different sizes of clusters. 
As shown in Table~\ref{tab:comparison}, {\sc{ReAssigner}} reduces ALW on CPU and memory on all the baselines for different cluster sizes.
{\sc{ReAssigner}} forces the PM to place requests more reasonably by pre-assigning the resources and adapting to the mismatch between the preset assigned resources and the incoming sequences by unassigning. 
Given $50$ PMs, {\sc{ReAssigner}} can reduce at most $54.72\%$ ALW on memory and $75.82\%$ ALW on CPU.
Given $100$ PMs, {\sc{ReAssigner}} can reduce at most $52.83\%$ ALW on memory and $70.01\%$ ALW on CPU.
Given $150$ PMs, {\sc{ReAssigner}} can reduce at most $52.21\%$ ALW on memory and $71.12\%$ ALW on CPU.

For the request length, {\sc{ReAssigner}} behaves slightly differently for different numbers of PMs in the cluster. 
In a cluster with $50$ PMs, the {\sc{ReAssigner}} improves $4.19\%$ mean request length for BF2, $2.06\%$ for Best-Fit, and $12.71\%$ for First-Fit.
It fails to do so with IB.
It is because VMs in scenarios with shorter lengths are more likely to come in extreme proportions, and sequences mismatch the pattern set by the assignment.
When more PMs are in the cluster (\ie, $100$ PMs), {\sc{ReAssigner}} improves by $6.99\%$ for BF2, $5.20\%$ for Best-Fit, $6.19\%$ for IB, and $14.70\%$ for First-Fit in the metric of average length. 
Given 150 PMs, {\sc{ReAssigner}} can improve by $5.51\%$ for BF2, $4.24\%$ for Best-Fit, $6.75\%$ for IB, and $14.42\%$ for First-Fit in the metric of average length.

Moreover, {\sc{ReAssigner}} achieves excellent improvement in scenes with longer processing sequences.
Given $100$ and $150$ PMs, {\sc{ReAssigner}} improves four baselines up to $18.28\%$ in the metric of top $50\%$ and top $75\%$ of the $60$ scenarios. Since the length of processing sequence given $50$ PMs is significantly less than $100$ PMs, the performance improvement of {\sc{ReAssigner}} for long sequences given $50$ PMs is only significant in FF and BF2.

\subsection{Ablation on unassigning strategy}

This section studies how two strategies of unassign influence our {\sc{ReAssigner}}'s performance. 
Unassign in {\sc{ReAssigner}} consists of two strategies: 
1) Unassign when imbalance (unassign under unbalanced utilization between virtual clusters): when the imbalance defined by (\ref{imbalance of two meta types}) exceeds the threshold $\alpha$, resources on a PM which has not placed any VM will be unassigned; 
2) Unassign when emergency (unassign when the emergent situation on allocation occurs): when the next VM cannot be placed on any PMs, {\sc{ReAssigner}} will search and choose the first PM that can handle the VM after unassigning.
We conduct an ablation experiment with IB given 100 PMs. 
The result is shown in Table~\ref{tab:ablation}.

Both unassignment improve the metric of average length but cause more ALW on resources.
The reduction of ALW on resources is due to the assignment because of the optimization (\ref{set_masking}), and the unassigning inevitably destroys this property by withdrawing the assignment of resources.
{\sc{ReAssigner}} makes a trade-off between balancing the usage of resources and processing more requests by the operation of unassignment so that the average length can be longer. The ALW on resources can be significantly reduced compared with the original algorithms.

\subsection{Ablation on assigning strategy}
This section considers impact of choosing different assigning strategy. 
The smallest CPU-Intensive request is 12U8G as shown in Table.\ref{tb: heter}.
Take it to perturb the CPU-Intensive assignment.
As shown in Table~\ref{tab:diff_mask}, the assignment means the assigned resources for CPU-intensive requests, which are 96U64G, 108U72G, 120U80G, 84U56G, respectively. 
Our assignment achieves the best performance on average, second and third quartile of the scheduling length, which is consistent with Section\ref{Performance comparison with baselines}.
It also can be noted that 84U56G achieves the smallest ALW, however it achieves significantly less scheduling length which makes it a poor choice of assignment.

\section{Conclusion}
\label{sec: conclusion}
This paper emphasizes that request heterogeneity impacts scheduling performance. 
Empirical results show that resource preference heterogeneity impacts scheduling performance critically. 
A plug-and-play intensifier, {\sc{ReAssigner}}, is proposed in this paper to improve the scheduling performances of the existing scheduler. 
{\sc{ReAssigner}} has three key sub-modules: {\em{categorize}}, {\em{assign}}, and {\em{unassign}}.
Through these modules, {\sc{ReAssigner}} generates multiple virtual clusters and lets each virtual cluster handle homogeneous requests only. 
Evaluating {\sc{ReAssigner}} in real traces of Huawei Cloud, {\sc{ReAssigner}} can intensify scheduling performances significantly for various existing schedulers.

\bibliographystyle{plainnat}
\bibliography{example}

\end{document}